\documentstyle[12pt]{report}
\begin{document}
\setlength{\baselineskip}{24pt}
\title{\large \bf COMPARISON OF THE SACHS-WOLFE EFFECT FOR \\
GAUSSIAN AND NON-GAUSSIAN FLUCTUATIONS}

\author{ \bf J. H. Kung \\
Harvard-Smithsonian Center for Astrophysics \\
60 Garden Street \\
Cambridge MA 02138 }

\date{}
\maketitle

\begin{abstract}

A consequence of non-Gaussian perturbations on the
Sachs-Wolfe effect is studied. For a particular power spectrum,
predicted Sachs-Wolfe effects are calculated for two
cases: Gaussian (random phase) configuration, and
a specific kind of non-Gaussian configuration.
We obtain a result that the Sachs-Wolfe effect for the latter
case is smaller when each temperature fluctuation is properly
normalized with respect to the corresponding mass
fluctuation ${\delta M\over M}(R)$.
The physical explanation and the generality
of the result are discussed.
\end{abstract}

\newpage

\begin{flushleft} {\large \bf I. INTRODUCTION}
\end{flushleft}

It is a challenging problem for cosmologists to explain the
large scale structures.  It is generally believed that some
microphysical processes  produced small amplitude perturbations
in the early universe, and that the present large scale structures
grew by gravitational instability of these perturbations.
Perturbations are called Gaussian (G) perturbations if the phases
of each fourier mode are random, and they are called non-Gaussian (NG)
perturbations if otherwise.

G initial fluctuations are well motivated in the
context of inflationary cosmology.  During the inflationary
phase, quantum fluctuations of a scalar field, which
drives the inflation, produces G fluctuations [1-4].
There have been many, both analytical and numerical, studies
of evolution of G fluctuations [5-7].
An evolutionary scenario requires further assumptions
about the type of matter in the universe, (e.g.,
Cold Dard Matter (CDM) or Hot Dark Matter (HDM)), and the initial
power spectrum of the fluctuations  $p(k) \equiv
{\vert \delta_k \vert}^2$.  In some cases even
$\Omega \equiv \rho /\rho_{critical}$ and
$\bf \wedge$, the cosmological constant, are left
as free parameters.

Statistical methods are typically used  to make comparisons
between predictions of a model and the observations.
Here a further assumption about the relationship between
the distribution of matter and the distribution of observed
galaxies needs to be made [8-10], i.e.,  biased galaxy
formation.

For a model to be branded a success it must fulfill at least
two requirements.  First, it must be capable of
reproducing the observed large scale structures.  Second, its
predicted Cosmic Microwave Background Radiation (CMBR) must
agree with the observations [11-14].  There are several mechanisms
contributing to the  CMBR temperature fluctuations, but
for fluctuations on  large angular scales ($\theta \gg 1^o$)
only two are important.  The first is the doppler effect
coming from our pecular velocity with respect to the
CMBR rest frame.  The second is the Sachs-Wolfe effect, which
is caused by  fluctuations in gravitational potentials on
the last scattering surface [15].
With the recent  results from the Cosmic Background
Explorer (COBE) [14], some G models are on the process of
being ruled out.  For example, HDM, $\Omega = 1$,
with the Harrison-Zel'dovich power spectrum predicts
too large ${\Delta T\over T}$ in direct
conflict with COBE [16,17].

An alternative to Gaussian models are whole class of
non-Gaussian (NG) models.  Examples of NG models are Cosmic
Strings, Monopoles, and Textures, which are the remnants of
phase transitions at early universe [18-23]. Production of
some NG fluctuations can even be accomodated within the
context of inflationary scenario [24-26].  There are also
mechanisms for generating NG perturbations without resorting
to 'exotic' physics of the early universe, for example the
explosion scenario of Ostriker and Cowie [27],
and Ikeuchi [28].

Detailed N body studies for evolution of NG perturbations
have been done [29-33], and their corresponding Sachs-Wolfe
effects were  calculated [34,35].  In this paper, we would like
to do a comparative study of Gaussian vs. non-Gaussian fluctuations
in the context of Sachs-Wolfe effect.
The  question
we would like to address is whether the   predicted
Sachs-Wolfe effect can be reduced by describing the
primordial density perturbations by a particular NG field
configuration
instead of a G field configuration.
That is, for a fixed power spectrum $p(k) = {\vert \delta_k\vert}^2$,
we would like to consider the consequences of two types of
field configurations, ${\delta \rho\over \rho}(\vec x)
= \sum \vert \delta_k\vert e^{i\phi (k)}e^{ik\cdot x}$.
The first is the G case where all the phases
$e^{i\phi(k)}$ are random.  The second is a NG case where
the phases are correlated in some way.  For each case we
will calculate the predicted ${\Delta T\over T}$ due to
the  Sachs-Wolfe effect and compare the magnitudes after
normalizing each ${\Delta T\over T}$ by the corresponding
mass fluctuation ${\delta M\over M}(R)$ at some scale $R$.

Even for a fixed power spectrum it is obvious that the set of
all possible NG field configurations is an infinite dimensional
parameter space.  There have been some analytical work
on the parametrization of NG perturbations ( Moscardini
et al. [36], Weinberg et al. [37]).  Because their
parametrization is in the 1 point probability function space
and not in fourier or configuration space their method is not directly
applicable here.  The reason is that every possible field
configuration described by a different set of phases $e^{i\phi (k)}$
gives a probability function
$P[\delta ]$, but each probability function does not give
a unique field configuration.  Therefore, in this
paper, we will first study the consequences of a very
special NG field configuration, i.e., $e^{i\phi (k)}$
are all +1 (or all -1).  Clearly the resulting configuration
${\delta \rho\over \rho}(\vec x)$
is a  spherical overdensity (spherical
void).  The difference between
this  NG and the G configuration is most accute at the
center of this object, where all the  amplitudes of different
wavelength  fluctuations add linearly in the NG case.

   In this paper, for this NG field configuration,
we have done a model
calculation for the power spectrum $p(k) \sim \delta (k - k_o)$.
As will be shown in detail, we arrive at the result that the
predicted Sachs-Wolfe effect for this NG case, when properly
normalized by mass fluctuation, is smaller than the
corresponding prediction for the G case.

   The layout of the paper is as follows. In section 2, some
notations and basic formuli are quoted from  literature.
In section 3,  ${\Delta T\over T}$  and ${\delta M\over M}(R)$
are calculated for the power spectrum $p(k) \sim \delta (k - k_o)$
as  G fluctuations. In sections 4 and 5, for the same
power spectrum, we calculate  ${\Delta T\over T}$ and
${\delta M\over M}(R)$ again, respectively, but now as
NG fluctuations. In section 6, the calculated temperature
fluctuations for the two cases are normalized with respect
to the corresponding mass fluctuation, and they are
compared.  In section 7, physical explanation of the
result is given and some comments and generalization are
given.

In the model calculations the scale factor of the universe
and the density perturbations were assumed to evolve as
in flat CDM universe.

\begin{flushleft}{\large \bf II.  PRELIMINARIES}
\end{flushleft}
\setcounter{chapter}{2}
\setcounter{equation}{0}

By definition, a spatial perturbation is just a fourier sum
of fluctuations on all wavelengths.
\begin{equation}
{\delta \rho\over \rho}(\vec x) = \sum_{\vec k}\vert \delta_{\vec k}\vert
e^{i\phi (\vec k)}e^{i{\vec k}\cdot {\vec x}}.
\end{equation}

\noindent
A power spectrum is defined as
\begin{equation}
p(\vec k) \equiv {\vert \delta_{\vec k}\vert }^2
\longrightarrow
 {\left( {L\over 2\pi}\right) }^3{\vert \delta_{\vec k}\vert }^2d^3k
{\rm \ continuum \ limit}.
\end{equation}

\noindent
$L$ is an arbitrary length which is large compare to any length
scale in the problem.
If the fluctuations were produced by some microphysical processes in the
early universe, then, a priori, there should be no preferred direction.
Therefore, rotational invariance dictates that $p(\vec k) = p(\vert k\vert)$.

\noindent
A model power spectrum we will use is

\begin{equation}
p(k) = \left\{ \begin{array}{ll}
\delta_o^2  &\vert k_o - {\Delta k\over 2}\vert \leq k
\leq \vert k_o + {\Delta k\over 2}\vert \\
 0  &{\rm otherwise.}
\end{array}
\right.
\end{equation}

\noindent
We will not use the Dirac delta function, i.e.,
$p(k) \sim \delta (k - k_o)$,
because for a NG case we will need to evaluate integrals of the
form $\int \vert \delta_k \vert d^3k\{ ... \}$, and square root
of a delta function is tricky to handle.  We will assume that  $\Delta k$
is thin enough so that $\int f(k)dk \approx f(k_o)\Delta k$ for any
reasonable function $f(k)$.

It turns out that this is the simplest power spectrum with which we
can demonstrate the essential differences between the G
and a particular NG configuration, which we will next discuss.

With this power spectrum, we will consider two different  field
configurations, i.e., choice of phases.

\begin{equation}
e^{i\phi (k)} = \left\{ \begin{array}{ll}
&{\rm  random, Gaussian \ case \ (G)} \\
&{\rm all +1 \ (or \  all -1), \ a \ non-Gaussian \ case \ (NG)}
\end{array} \right.
\end{equation}

The physics behind generating a CMBR fluctuation from a density
fluctuation is easy to understand.  Simply stated, a matter
perturbation distorts space, and therefore it 'distorts'
geodesics of photons which were last scattered at recombination.
Because a Newtonian potential is time independent in  a linear
regime only the perturbation in the
Newtonian potential on the last scattering surface is important
in the linear Sachs-Wolfe calculation. Contributions
from any intermediate distortion  of a geodesic is unimportant
by the blue and redshift cancellation effect.
If a structure becomes nonlinear, then the gravitational potential
is no longer constant, and the cancellation is not complete.
This higher order effect has been studied by Rees and Sciama [38].

The relationship between a matter perturbation and ${\Delta T\over T}$
can be described most succintly by the expression [15, 39]

\begin{equation}
{\Delta T\over T}(\hat q, \vec x) \equiv
{ T(\hat q, \vec x) - T_o\over T_o} =
{1\over 3}
\left[ \delta\phi(\vec x + \tau_o\hat q) -
 \delta\phi(\vec x) \right],
\end{equation}

\noindent
where $ \delta\phi(\vec x)$ is the Newtonian potential
at point $\vec x$,  $\tau_o = 2H_o^{-1}$, $\hat q$ is a direction
in sky, and $T_o = 2.7 K$ is the mean CMBR temperature.
{}From $\nabla^2\delta\phi = 4\pi G\delta \rho$, we get

\begin{equation}
\delta\phi(\vec x)= -{3\over 2}H_o^2\sum_k\delta_k(t_o)k^{-2}
e^{i\vec k\cdot \vec x}.
\end{equation}

\noindent
Rigorously, these expressions are valid in time-orthogonal coordinate
and for a flat universe.  The generalization to an open or closed
universe is discussed by Anila and Motta [40].

\begin{flushleft}{\large \bf III. GAUSSIAN FLUCTUATIONS}
\end{flushleft}
\setcounter{chapter}{3}
\setcounter{equation}{0}

For G fluctuations, because of the random phase of each
$\delta_k$ (2.1), only statistical information can be studied,
e.g., the angular correlation function of  ${\Delta T\over T}$.
Using (2.5) and (2.6) we get

\begin{eqnarray}
C(\hat q_1 \cdot \hat q_2)
&\equiv
 &\langle {\Delta T\over T}(\hat q_1,\vec x)
 {\Delta T\over T}(\hat q_2,\vec x)\rangle_{space} \nonumber \\
&=
&\pi H_{o}^{4}{\left( {L\over 2\pi }\right)^3}
 \int {dk\over k^2}p(k) \left( j_o(k\tau_o \vert \hat q_1 - \hat q_2 \vert )
 - {\rm monopole - dipole} \right).
\end{eqnarray}

\noindent
The monopole (i.e., spherically symmetric) term
is  subtracted out because it is unobservable for
a fixed observer.  The dipole term is also subtracted out because
in principal it is impossible to distinguish an intrinsic dipole
anisotropy from the observer's pecular velocity.
The monopole and the dipole terms can be identified by using the
following identity, which is proven in the appendix.

\begin{equation}
j_o(\vert A\hat A - B\hat B\vert ) = \sum_l(2l+1)j_l(A)
j_l(B)p_l(\hat A\cdot \hat B)
\end{equation}

\noindent
$j_o$ and $p_l$ are Spherical Bessel  and Legendre functions,
respectively.
Setting $A=k\tau_o$, $\hat A=\hat q_1$, $B=k\tau_o$, and
$\hat B=\hat q_2$ we identify the monopole and the dipole
terms to be $ j_o^2(k\tau_o)$ and $3j_1(k\tau_o)^2p_1(\hat q_1\cdot
\hat q_2)$, respectively.

\noindent
Using a thin $\Delta k$ approximation to evaluate the integral
for the power spectrum (2.3), we finally get [15,41]

\begin{equation}
C(\hat q_1 \cdot \hat q_2)
= \pi H_o^4{\left( {L\over 2\pi }\right)}^3
  {\Delta k\over k_o^2}\delta_o^2 \left( {\sin {\varsigma k_o}\over
\varsigma k_o} - j_o^2(k_o\tau_o) - 3j_1(k_o\tau_o)^2p_1(\hat q_1 \cdot
\hat q_2)\right),
\end{equation}

\noindent
with $\varsigma \equiv \tau_o\vert \hat q_1 - \hat q_2 \vert$.
The overall normalization $\delta_o$ must be fixed. This is usually
done  by relating it to
either ${\delta M\over M}(R)$ or mass autocorrelation function
$\xi (\vec r) \equiv \langle {\delta \rho\over \rho}(\vec x + \vec r)
 {\delta \rho \over \rho}(\vec x)\rangle_{space}.$
Normalizing $\delta_o$ by $\xi (\vec r)$ would be subjecting
the model to more stringent requirements, which would be appropriate
for a more realistic power spectrum like $p(k) \sim k^n $.
Therefore, for our simplistic models, we choose to normalize it by
 ${\delta M\over M}(R)$.

Again for the G case, we will just quote the well known result
relating the power spectrum and  ${\delta M\over M}(R)$ [39].

\begin{equation}
\langle \vert {\delta M\over M}(R)\vert^2\rangle_{space} =
{\left( {L\over 2\pi }\right)}^3
 \int d^3k p(k) W(kR)
\end{equation}

\noindent
where $W(kR)$ is a window function.  We will use the top
down window function

\begin{equation}
W(y) = 9y^{-6}\left( \sin y - y\cos y\right)^2.
\end{equation}

\noindent
Evaluating the expression for the power spectrum (2.3) we get

\begin{equation}
\langle \vert {\delta M\over M}(R)\vert^2 \rangle_{space} =
4\pi {\left( {L\over 2\pi }\right)}^3
 k_o^2\Delta k \delta_o^2 W(k_oR).
\end{equation}

\begin{flushleft}{\large \bf IV.
${\Delta T\over T} $ FOR NON-GAUSSIAN FLUCTUATIONS}
\end{flushleft}
\setcounter{chapter}{4}
\setcounter{equation}{0}

In this section, we would like to recalculate the temperature
fluctuation for the same power spectrum (2.3) but now for the
NG configuration (2.4).
Recall that in the G case the best one can do was to evaluate
{\it rms} of various quanties of interest.  For a NG case this is
no longer necessary.  For NG perturbations it is possible to get an
explicit expression for a temperature fluctuation as a function
of  position $\vec x$ in a direction $\hat q$.
Using the fact that $e^{i\phi (k)} = 1$, it is trivial to
do the angular sum  $d\Omega_k$ in Eq. (2.6).

\begin{equation}
{\Delta T\over T}(\hat q, \vec x,\tau_o)
= -2\pi H_{o}^{2}{\left( {L\over 2\pi }\right)}^3
 \int dk p(k)^{1/2}
\left( j_o(k\vert \vec x + \tau_o\hat q\vert ) - {\rm monopole - dipole}
 \right).
\end{equation}

\noindent
Using a thin $\Delta k$ approximation and
Eq. (3.2) with  $A = k\tau_o$, $\hat A = \hat q$,
$B = kr$,  and $\hat B = \hat x$, to identify the monopole and
the dipole terms, we get

\begin{equation}
{\Delta T\over T}(\hat q, \vec x,\tau_o)
= -2\pi H_{o}^{2}{\left( {L\over 2\pi }\right)}^3
\Delta k_o \delta_o
\left( \begin{array}{l}
{\sin {\varsigma k_o}\over \varsigma k_o}  - j_o(k_o\tau_o)j_o(k_or) \\
 + 3j_1(k_o\tau_o)j_1(k_or)p_1(\hat q\cdot \hat x)
\end{array} \right),
\end{equation}

\noindent
where now $\varsigma \equiv \vert \tau_o \hat q + r\hat x \vert $.
It should be noted that this is the result for  the specific
NG configuration (2.4), and other NG configurations will give  different
results.

\begin{flushleft}{\large \bf V.
${\delta M\over M}(R)$ FOR NON-GAUSSIAN FLUCTUATIONS}
\end{flushleft}
\setcounter{chapter}{5}
\setcounter{equation}{0}
{}

We would now like to  derive a simple expression for a
mass fluctuation at some scale R that is appropriate for
NG fluctuations.  Again, for a NG perturbation it is no longer
necessary to evaluate {\it rms }of quantities of interest.  By definition
(2.1), for (2.4)
\begin{equation}
{\delta \rho\over \rho }(\vec x) =  {\left( {L\over 2\pi }\right)}^3
 \int k^2dk d\Omega_k \vert \delta_k\vert
\left[ {e^{i\phi (k)}\over 1}\right]e^{i\vec k\cdot \vec x}.
\end{equation}

\noindent
The angular integral $d\Omega_k$ is simple to evaluate.

\begin{equation}
{\delta \rho\over \rho }(r) = 4\pi {\left( {L\over 2\pi }\right)}^3
 \int k^2dk  \vert \delta_k\vert
{\sin (kr)\over kr}.
\end{equation}

\noindent
Defining ${\delta M\over M}(R)$ as an excess mass within a sphere
of radius $R$ centered on a point where the density contrast is
maximum, we get

\begin{equation}
{\delta M\over M}(R) =
4\pi {\left( {L\over 2\pi }\right)}^3
 \int k^2dk \vert \delta_k \vert \sqrt{ W(kR)},
\end{equation}

\noindent
where $W(kR)$ is the familiar top down window function (3.5).
Evaluating this for power spectrum (2.3) we get

\begin{equation}
{\delta M\over M}(R) =
4\pi {\left( {L\over 2\pi }\right)}^3
  k_o^2\Delta k \delta_o \sqrt{ W(k_oR)}.
\end{equation}

\begin{flushleft}{\large \bf VI.
COMPARISON OF TWO TYPES OF PERTURBATIONS}
\end{flushleft}
\setcounter{chapter}{6}
\setcounter{equation}{0}
{}

We will now use the simplest method to compare the magnitudes
of two ${\Delta T\over T}$ as predicted by each model.
We will first write each  ${\Delta T\over T}^{\rm NG(G)}$
in terms of the corresponding
${\delta M\over M}(R)^{\rm NG(G)}$.  Then we will compare the
magnitudes of  ${\Delta T\over T}^{\rm NG}$ and
 ${\Delta T\over T}^{\rm G}$ by setting
${\delta M\over M}(R)^{\rm NG} = {\delta M\over M}(R)^{\rm G}$.
Using (3.3) with (3.6), we get for the G case

\begin{equation}
\left ( {\Delta T\over T}\right)_{rms}^G(\hat q_1, \hat q_2)
= {1\over 2}W(k_oR)^{-1/2}H_o^2k_o^{-2}
\left( {\delta M\over M}(R)\right)_{rms}^G
 \left\vert {\sin {\varsigma k_o}\over
\varsigma k_o} - j_o^2(k_o\tau_o) - 3j_1(k_o\tau_o)^2p_1(\hat q_1 \cdot
\hat q_2)\right\vert^{1/2}
\end{equation}

\noindent
with $\varsigma = \tau_o\vert \hat q_1 -  \hat q_2\vert$.
For the NG case using (4.2) with (5.4),  we get

\begin{equation}
{\Delta T\over T}^{NG}(\hat q, \vec x)
= - {1\over 2}W(k_oR)^{-1/2}H_o^2k_o^{-2}{\delta M\over M}^{NG}(R)
\left( \begin{array}{l}
{\sin {\varsigma k_o}\over \varsigma k_o}  - j_o(k_o\tau_o)j_o(k_or) \\
 + 3j_1(k_o\tau_o)j_1(k_or)p_1(\hat q\cdot \hat x)
\end{array} \right)
\end{equation}

\noindent
with $\varsigma = \vert \tau_o\hat q + r\hat x\vert$.
The graph of  ${\Delta T\over T}$ for G case is shown in
Figure 1.  The graphs of  ${\Delta T\over T}$ for
the NG case are shown in Figures 2(a)-2(f) for various
values of $k_or$.  It is interesting to note that for the NG
case the temperature anisotropy depends on position
in space.  This is because the information about the positions
of mass perturbations is not lost in the ${\Delta T\over T}$
calculation for a  NG case, whereas in the corresponding calculation for
the G case they are lost by incorporating the {\it rms} methods.

It should be noted that  we are comparing the magnitudes
of {\it rms} of  autocorrelation function for ${\Delta T\over T}$
with the local temperature fluctuation; therefore,
the  angular parameter $\theta$ that appears
in the figures for both G and NG have  different meanings.

In graphing  ${\Delta T\over T}'s$, a common normalization
was used for both G and NG cases.  From the figures we
can conclude that, for the  models described by (2.3) and (2.4),
the predicted
Sachs-Wolfe effect for the NG case is about
$ 3\cdot 10^{-3}$ smaller than for the G case.  Actually this
conclusion is true as long as $\vert \tau_o\hat q - r\hat x \vert
\gg k_o^{-1}$.  In other words,
the position  of  spherical overdensity (void)
does not intersect the locus of
last scattering surface in relation to the observer
at $\vec x$.

\begin{flushleft}{\large \bf VII.  DISCUSSION}
\end{flushleft}
\setcounter{chapter}{7}
\setcounter{equation}{0}

Before getting into the discussion of the generality of the
result, some physical explanations are in order.
First, the  difference between G and the NG
configuration (2.4) on ${\delta \rho\over \rho}(\vec x)$ is
obvious.  For the NG case, there are regions in space where
the amplitudes of different wavelength fluctuations
are adding coherently.  In these regions
the magnitudes of ${\delta \rho\over \rho}(\vec x)$
are larger than the {\it rms} of ${\delta \rho\over \rho}(\vec x)$,
which is a relevant quantity for the G case. With this
understanding, the conclusion that the Sachs-Wolfe effect
for the NG configuration (2.4) is smaller than the Sachs-Wolfe
effect for the G configuration may seems contrary to Eq. (2.5).
After all, if matter is more 'clumpy' in the NG configuration (2.4),
than the maximum difference in the Newtonian potential should
be larger for the NG case. And therefore, ${\Delta T\over T}$
for the NG configuration should also be larger.  The loophole in this
 argument lies in two facts.  First, (2.5) needs to be normalized
because of the quantity $\delta_o$ (see (2.3), (3.3), (4.2)).
Second, in calculating
the Sachs-Wolfe effect $l = 0$ (monopole) term of
${\delta T\over T}$ is subtracted out because it is
unobservable.  For the NG configuration (2.4) under consideration,
it is  easy to see that regions in space where
${\Delta T\over T}(\hat q, \vec x)$ is much larger than the {\it rms}
of ${\Delta T\over T}(\hat q, \vec x)$ are spherically symmetric.
Therefore, when the $l=0$ term is subtracted out the
enhancement of going from a linear sum to the 'random walk'
sum largely disappears.

We would now like to give a proof that  the NG configuration
(2.4) among all possible NG configurations
gives the smallest  Sachs-Wolfe effect for the power spectrum (2.3).
As stated in the introduction, even for a fixed power spectrum,
the set of all NG field configurations is an infinite parameter space.
For the argument we choose to parametrize a  NG configuration
in a following way.  A general phase of a fluctuation
$\delta_k$ can be expanded in terms of spherical harmonics

\begin{equation}
e^{i\phi (\vec k)} = \sum_{l,m}Y_{l,m}(\hat k)G_{l,m}(\vert k\vert ).
\end{equation}

\noindent
For the power spectrum (2.3) we are considering, $\vert k\vert$ is
constant, so $G_{l,m}(k)$ are just constants. Since
$e^{i\phi }$ is a unit norm, it implies that
${\sum \vert G_{l,m}\vert }^2 = 4\pi$  or $ \vert G_{l,m}\vert \leq
\sqrt{4\pi}$.
In this general NG field configuration, it is easy to show that
for fixed power spectrum the
new mass fluctuation is (Eqs. (5.1)-(5.4))

\begin{equation}
\left. {\delta M\over M}(R)\right\vert_{NG}^{new} =
{G_{o,o}\over \sqrt{4\pi }}
\left.{\delta M\over M}(R)\right\vert_{NG}^{old}.
\end{equation}

\noindent
By old we mean the previously discussed case $e^{i\phi} = 1$.
We are being little cavalier here.  Even though the
resulting density configuration ${\delta \rho\over \rho }(\vec x)$
may not be spherically symmetric, we are  defining
$ {\delta M\over M}(R)$ as an excess mass
within a  sphere of radius $R$ centered on  a point
where the density contrast is maximum.

The change in the Newtonian potential is more complicated.
Inserting (7.1) into (2.6) and after $d\Omega_k$ we get

\begin{equation}
\delta\phi(\vec x + \tau_o \hat q)
= -6\pi H_{o}^{2}{\left( {L\over 2\pi }\right)}^3
\Delta k \delta_o
\sum_{l,m}i^lY_{l,m}(\hat \varsigma)j_l(k_o\varsigma)G_{l,m}
\end{equation}

\noindent
with $\vec \varsigma \equiv \vec x + \tau_o\hat q$,  and
${\Delta T\over T}(\hat q, \vec x) = {1\over 3}[\delta \phi(\vec \varsigma ) -
{\rm monopole}]$ with (see (2.5))

\begin{equation}
{\rm monopole \ term } =
{\int \delta\phi(\vec \varsigma)d\Omega_{\hat q}\over
\int d\Omega_{\hat q}} = \delta \phi(\vec x)j_o(k_o r).
\end{equation}

\noindent
Denoting  quantities with the $\delta_o$ and $G_{l,m}$ dependence
extracted by a hat (e.g., $\hat A$) we get from (7.3), (7.4),
and (2.5)

\begin{equation}
{\Delta T\over T}(\hat q, \vec x) = {\delta_o\over 3}\sum_{l,m} G_{l,m}\left[
 \widehat{\delta\phi}_{l,m}(\vec \varsigma) -
 \widehat{\delta\phi}_{l,m}(\vec x)j_o(k_or)
 \right].
\end{equation}

\noindent
And from (5.4) and (7.2) we get
${\delta M\over M} = \delta_o G_{oo}{\widehat{\delta M}\over M}$.
Eliminating $\delta_o$, we need to find the extremum of

\begin{equation}
{\Delta T\over T}(\hat q, \vec x) = {G_{oo}^{-1}\over 3}
{\delta M\over M}({\widehat{\delta M}\over M})^{-1}
\sum_{l,m} G_{l,m}\left[
 \widehat{\delta\phi}_{l,m}(\vec \varsigma) -
 \widehat{\delta\phi}_{l,m}(\vec x)j_o(k_or)
\right]
\end{equation}

\noindent
subject to the constraint ${\sum \vert G_{l,m}\vert }^2 = 4\pi$.
With the method of lagrange multiplier, we get
${\sum \vert G_{l,m}\vert }^2 = G_{oo}^2$, or
$e^{i\phi} = G_{oo}Y_{oo} = \pm 1$.  It is easy to check that the
extremum is a minimum because another configuration, e.g.,  $G_{oo} \approx 0$,
gives a larger value.  It should be noted that the variational
method alone does not tell us that the field configuration
$e^{i\phi} = \pm 1$ gives a Sachs-Wolfe effect that is
smaller than for the G configuration.
A remark on the measure of generality of the result is
possible.  ${\Delta T\over T }$ given by  Eq. (7.6)
 is a smooth linear function of
$G_{l,m}$ except for $G_{oo}$.  And therefore, combined with the restriction
of ${\sum \vert G_{l,m}\vert }^2 = 4\pi$,
the conclusion will be unchanged
as long as $G_{oo} \approx \pm \sqrt{4\pi}$.

The result for a more realistic power spectrum
$p(k) \sim k^n$ will be discussed elsewhere [42].  The decrease in the
Sachs-Wolfe effect persists but to a different degree.

In light of the result that there are NG field configurations that can reduce
the Sachs-Wolfe effect, we could turn the analysis on its head.
By properly incorporting void evolution, we could have addressed
the question of the largest possible void compatible with COBE
given a  NG density fluctuations
(here $e^{i\phi} \approx -1$).
This would be the NG counterpart of the work by Blumenthal et al. [43].

And finally, even though in our analysis we have assumed that
perturbations and the scale factor of the universe evolved
as that of flat FRW universe, the decrease in the Sachs-Wolfe
effect will obviously persist in other cosmologies but to
different degrees.

\begin{flushleft}{\large \bf IIX.  CONCLUSION}
\end{flushleft}
\setcounter{chapter}{8}

In this paper, we have taken a power spectrum, which was
essentially a delta function,  and have shown that there is
a set of non-Gaussian field configurations for which the Sachs-Wolfe
effect is smaller than  the corresponding effect for the
Gaussian configuration
case.  This set of non-Gaussian perturbations is characterized
by having a large monopole term , $G_{oo} \approx \pm \sqrt{4\pi}$,
in the $Y_{l,m}$ decomposition of the phases, i.e.,
$e^{i\phi (\vec k)} =
\sum_{l,m}Y_{l,m}(\hat k)G_{l,m}(\vert k\vert ) \approx \pm 1 .$

The conflict between the recent COBE measurements of CMBR
and some HDM Gaussian models can potentially
be salvaged by the combination of a non-Gaussian fluctuation model
and HDM.

\begin{center}
{\large \bf APPENDIX}
\end{center}

We would like to prove an identity.

$$
j_o(\vert A\hat A - B\hat B\vert ) = \sum_l(2l+1)j_l(A)
j_l(B)p_l(\hat A\cdot \hat B)
$$

\noindent
consider

\begin{eqnarray*}
\int e^{i(A\hat A - B\hat B)\cdot \hat C}d\Omega_c
&=& \int\sum_l i^l(2l+1)j_l(\vert A\hat A - B\hat B\vert)
 p_l((A\hat A - B\hat B)\cdot \hat C)d\Omega_c \\
&=& 4\pi j_o(\vert A\hat A - B\hat B\vert ) \\
&=& \int\sum_{l,l'}\left[ i^l(2l+1)j_l(A)
 p_l(\hat A\cdot \hat C)\right]
\left[ (-i)^{l'}(2l'+1)j_{l'}(B)
 p_{l'}(\hat B\cdot \hat C)\right] d\Omega_c    \\
& &  {\rm using} \int p_l(\hat A\cdot \hat C)
p_{l'}(\hat B\cdot \hat C)d\Omega_c = {4\pi \over 2l+1}
p_l(\hat A\cdot \hat B)\delta_{l l'} \\
&=& 4\pi \sum_l(2l+1)j_l(A)j_l(B)p_l(\hat A\cdot \hat B).
\ \ Q.E.D \end{eqnarray*}

\begin{center}
{\large \bf ACKNOWLEDGEMENTS}
\end{center}

I would like to thank George Field, T. Piran, G. Efstathiou,
Hardy Hodges, Helmut Zaglauer, for discussions
and valuable comments, Dalia Goldwirth for bring
to my attention a pedagogical review of Sachs-Wolfe effect
by G. Efstathious, and R. Brandenberger for carefully reading
the original manuscript and for his many constructive
criticisms.

\newpage
\begin{center}
{\large \bf FIGURE CAPTIONS}
\end{center}
\begin{enumerate}
\item Graph of the square root of the temperature autocorrelation
function
$\vert \left< {\Delta T(\theta_o)\over T}
 {\Delta T(\theta_o +\theta)\over T}\right>\vert^{1/2}$,
Eq. (6.1), as function of $\theta$.
$H^{-1} = 3000MPC$ and $2\pi k_o^{-1} = 40MPC$ was used to
imitate the peak in the power spectrum for a HDM.
The normalization is arbitrary.
\item Graph of the temperature fluctuation
${\Delta T\over T}(\hat q, \vec x)$, Eq. (6.2), as function of
$\theta=cos^{-1}(\hat q\cdot ~\hat x)$ for
various values of $r = \vert \vec x \vert$.
Figures 2(a)-2(f) correspond to $k_or = [0, 2.0, 4.0, 6.0, 8.0, 10.0]$,
respectively. The values of $H_o$, $k_o$, and the overall
normalization are the same as those of Fig. 1.
\end{enumerate}

\newpage
\begin{center}
{\large \bf REFERENCES}
\end{center}

\begin{enumerate}
\item A. H. Guth, and S. Y. Pi, Phys. Rev. Lett. {\bf 49}, 1110 (1982).
\item S. W. Hawking, Phys. Lett. {\bf B 115}, 295 (1982).
\item A. A. Starobinsky, Phys. Lett. {\bf B 117}, 175 (1982).
\item J. M. Bardeen, P. J. Steinhardt, and M. S. Turner, Phys. Rev. {\bf D 28},
679 (1983).
\item G. Efstathiou, C. S. Frenk, S. D. M. White, and M. Davis, Mon. Not.
R. Astron. Soc. {\bf 235}, 715 (1988).
\item T. Suginohara, and Y. Suto, Astrophys. J. {\bf 371}, 470 (1990).
\item J. F. Beacom, et al., Astrophys. J. {\bf 372}, 351 (1991).
\item M. Davis, G. Efsthathious, C. S. Frenk, and S. D. M. White,
 Astrophys. J. {\bf 292}, 371 (1985).
\item C. Park, Mon. Not. R. Astron. Soc. {\bf 251}, 167 (1991).
\item N. Kaiser, Astrophys. J. {\bf 284}, L9 (1985).
\item P. Meinhold, and P. Lubin, Astrophys. J. {\bf 370}, L11
(1991).
\item S. S. Meyer, E. S. Cheng, and L. A. Page,
Astrophys. J. {\bf 371}, L7 (1991).
\item G. F. Smoot, et al., Astrophys. J. {\bf 371}, L1, (1991).
\item G. F. Smoot, et al., Astrophys. J. Lett. (submitted).
\item R. K. Sachs, and A. M. Wolfe, Astrophys. J. {\bf 147}, 73
(1967).
\item E. W. Kolb, and M.S. Turner,  {\it The Early Universe}
(Addison-Wesley Publishing Company, 1990).
\item E. L. Wright, et al., Astrophys. J. Lett. (submitted).
\item T. W. B. Kibble, J. Phys. {\bf A9}, 1387 (1976).
\item Ya. B. Zel'dovich, Mon. Not. R. Astro. Soc. {\bf 192},
663 (1980).
\item A. Vilenkin, Phys. Rep. {\bf 121}, 263 (1985).
\item N. Turok, Phys. Rev. Lett. {\bf 63}, 262 (1989).
\item M. Barriola, and A. Vilenkin, Phys. Rev. Lett.
{\bf 63}, 341 (1989).
\item D. P. Bennett, and S. H. Rhie, Phys. Rev. Lett.
{\bf 65}, 1709 (1990).
\item T. J. Allen, B. Grinstein, and M. B. Wise, Phys. Lett.
{\bf B 197}, 66 (1987).
\item D. S. Salopek, J. R. Bond, and J. M. Bardeen,
Phys. Rev. {\bf D 40}, 6 (1989).
\item L. A. Kofman, and A. D. Linde, Nucl. Phys. {\bf B 282},
555 (1987).
\item J. P. Ostriker, and L. L. Cowie, Astrophys. J. {\bf 243},
L127 (1981).
\item S. Ikeuchi, Publ. Astron. Soc. Japan {\bf 33}, 211 (1981).
\item R. J. Scherrer, A. L. Melott, and E. Bertschinger,
Phys. Rev. Lett. {\bf 62}, 379 (1989).
\item C. Park, D. N. Spergel, and N. Turok, Astrophys. J.
{\bf 372}, L53 (1991).
\item A. K. Gooding, C. Park, D. N. Spergel, N. Turok, J. R. Gott,
Astrophys. J. (in press).
\item R. Cen, J. P. Ostriker, D. N. Spergel, N. Turok,
Astrophys. J. (in press).
\item M. J. West, D. H. Weinberg, A. Dekel, Astrophys. J.
{\bf 353}, 329 (1990).
\item J. Traschen, N. Turok, and R. Brandenberger,
Phys. Rev. {\bf D 34}, 919 (1986).
\item N. Turok, and D. N. Spergel, Phys. Rev. Lett. {\bf 64},
2736 (1990).
\item L. Moscardini, S. Matarrese, F. Lucchin, and A. Messina,
Mon. Not. R. Astro. Soc. {\bf 248}, 424 (1991).
\item D. H. Weinberg, and S. Cole, Mon. Not. R. Astro. Soc.
(submitted).
\item M. J. Rees, and D. W. Sciama, Nature, {\bf 217}, 511 (1968).
\item P. J. E. Peebles, {\it The Large Scale Structure of the Universe}
(Princeton University Press, Princeton, 1980).
\item A. M. Anile, and S. Motta, Astrophys. J. {\bf 207}, 685 (1976).
\item E. Martinez-Gonzales, and J. Sanz, Astrophys. J. {\bf 347},
11 (1989).
\item J. H. Kung, (in preparation).
\item G. R. Blumenthal, et al., Astrophys. J. {\bf 338},
225 (1992).
\end{enumerate}
\end{document}